\documentclass[letterpaper, 11pt, onecolumn]{IEEEtran}
\usepackage{amsmath,amssymb,amsfonts}
\usepackage{algorithmic}
\usepackage{graphicx,subfigure}
\usepackage{textcomp}
\usepackage{xcolor}
\usepackage{booktabs}
\usepackage{multirow}
\usepackage[T1]{fontenc}
\usepackage{soul}
\usepackage{changepage,threeparttable}

\usepackage{cite}
\usepackage{caption}
\usepackage[font=small,labelfont=bf]{caption}
\begin{document}

%\title{A New Accurate Non-Invasive EEG-Based Schizophrenia Detection System in IoMT Framework}

\title{Novel EEG based Schizophrenia Detection with IoMT Framework for Smart Healthcare}

\author
{
\IEEEauthorblockN{Geetanjali Sharma\IEEEauthorrefmark{1}\IEEEauthorrefmark{2},
Amit M. Joshi\IEEEauthorrefmark{1},\\
\IEEEauthorblockA{\IEEEauthorrefmark{1}Malaviya National Institute of Technology, Department of Electronics and Communication, Jaipur, India}\\
\IEEEauthorblockA{\IEEEauthorrefmark{2}Maharaja Surajmal Institute of Technology, Department of Electronics and Communication, Delhi, India}}
}

\maketitle

\begin{abstract}
In the field of neuroscience, Brain activity analysis is always considered as an important area. Schizophrenia(Sz) is a brain disorder that severely affects the thinking, behavior, and feelings of people all around the world.  Electroencephalography (EEG) is proved to be an efficient biomarker in Sz detection.
EEG is a non-linear time-series signal and utilizing it for investigation is rather crucial due to its non-linear structure.  This paper aims to improve the performance of EEG based Sz detection using a deep learning approach. A novel hybrid deep learning model known as SzHNN (Schizophrenia Hybrid Neural Network), a combination of Convolutional Neural Networks (CNN) and Long Short-Term Memory (LSTM) has been proposed. CNN network is used for local feature extraction and LSTM has been utilized for classification. The proposed model has been compared with CNN only, LSTM only, and machine learning based models. All the models have been evaluated on two different datasets wherein Dataset 1 consists of 19 subjects and Dataset 2 consists of 16 subjects. Several experiments have been conducted for the same using various parametric settings on different frequency bands and using different sets of electrodes on the scalp. Based on all the experiments, it is evident that the proposed hybrid model (SzHNN) provides the highest classification accuracy of 99.9\% in comparison to other existing models. The proposed model overcomes the influence of different frequency bands and even showed a much better accuracy of 91\% with only 5 electrodes. The proposed model is also evaluated in Internet of Medical Things (IoMT) framework for smart healthcare and remote monitoring applications.
\end{abstract}


%\end{frontmatter}

% \linenumbers

%%%%%%%%%%%%%%%%%%%%%%%%%%%%%%%%%%%%%%%%%%%

\section{Introduction}

%\textcolor{red}{Title contains Scz as Schizophrenia make that Uniform. Draw the thematic Diagram in Section I to show overview of your approach. Divide Section I first long paragraph in two paragraph, Please consider to rewrite the sentence with all blue lines of the paper}
%\textcolor{green}{all done, and blue line sentences are also improved and highlighted in green}

\begin{figure}[!h]
\centerline{\includegraphics[scale=0.4]{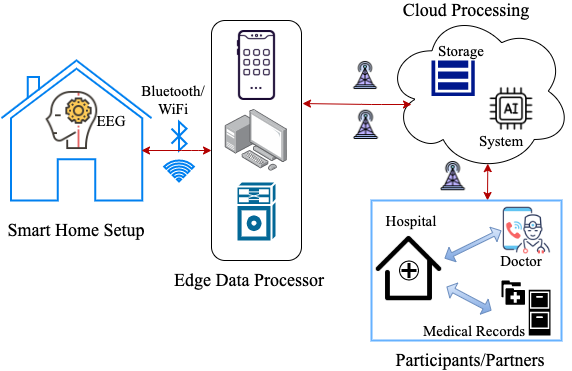}}
\caption{IoMT-based smart healthcare framework for Sz detection.}
\label{fig:frame}
\end{figure}

\IEEEPARstart{D}{iseases}  are often diagnosed in the medical field utilizing imaging modalities, laboratory testing, and biological markers. Though, the diagnosis of diseases 
involving psychiatric instability is 
mainly based on patient interviews, symptoms explained, and the presence or absence of 
representative signs 
of attitude and behavior. Schizophrenia (Sz) is a serious neurological disorder that affects people in several parts of their normal thinking, emotions, and behavioral characteristics \cite{wing1985management}. 
The World Health Organization(WHO) reports that Sz affects more than 21 million individuals worldwide. 
Moreover, The National Institute of Mental Health reports Sz as a major contributor to disease burden, reporting that 2.4 million adults over the age of 18 are affected by it in the United States only. Symptoms of Sz can include hearing non-existing voices, hallucinations, 
disorganized speech, and response depletion, and many more \cite{friston1995schizophrenia}. Sz is extremely genetic, people with both the Zinc Finger protein 804A and genes Neurogranin are at a higher risk of developing it \cite{williams2011fine}. 
Quality of life is also significantly impacted, as the majority of Sz patients are being unable to work and around 20-40\% would attempt for suicide. Therefore, an accurate and rapid pre-diagnostic method is very much required for proper care and quality treatment of the patients \cite{joshi2020iglu2.0}. 
In Sz, standard methods of the identification of at-risk people based on an analysis of family 
history with disease to target individuals who are 
in the premorbid and/or initial stages of the disorder\cite{savio2010neural}.\\
The healthcare system has seen rapid growth from conventional to mobile health, e-health and, smart healthcare with the use of the Internet of Medical Things (IoT) \cite{jain2020iglu}. IoMT helps to have the real-time solutions of the healthcare problems with remote access of patient data \cite{jain2019iomt}.  The EEG data can be captured from the patients continuously using the wearable device. The data can be stored at cloud for long term observation by health experts who would suggest proper point of care solution \cite{jain2019iglu1.0}. An overview of the proposed smart connected healthcare framework with Sz detection is shown in the Figure \ref{fig:frame}. 
The sz is detected using machine learning algorithm which would help to have proper treatment of the patient. 
This proposes the requirement of a system having good feature extraction and classification with lesser computational cost \cite{pancholi2019novel}.
This paper proposes a computer-aided hybrid deep neural network based detection system utilizing EEG signals as the biomarker for effective diagnosis of Sz disorder in smart healthcare environment.
There are mainly 4 components of this framework, 1) EEG Data Acquisition 2) Edge data processor 3) Cloud storage and, 4) control system. In the present case, there would be smart healthcare solution  wherein EEG being captured then gets transferred to Edge data processor with the help of Bluetooth or Wi-Fi. Here, the edge data processor pre-processes the EEG to reduce complexity, so that data can be easily transferred and stored on the cloud with minimum bandwidth. Thereafter, feature extraction and classification of EEG data for Sz detection is done with the help of the proposed automated Sz detection model on the cloud itself. The weight parameters are being stored on the cloud for further processing. The health records of the patient may accessed by the doctor and/or hospital or the healthcare provider who can prescribe the necessary treatment to the patient, if required \cite{joshi2020iglu2.0}. The necessary feedback could be provided with control system to patient through IoMT based framework for sz detection \cite{joshi2020secure}.\\
Till now, diagnosis of Sz completely depends on behavioral markers found by the specialists due to the unavailability of any standard clinical test \cite{van2014brain},\cite{sharma2021automated}. These types of assessments are not precise they can’t detect abnormalities within the brain.  So an automated computer-aided solution with very high accuracy and less complexity can help in the effective diagnosis of Sz.
Up until now, many non-invasive methods have been implemented and proposed by researchers to detect Sz based on the EEG signals and imaging techniques \cite{akar2016analysis},\cite{krizhevsky2012imagenet}. However, unlike EEG the imaging methods such as Computed Tomography and Medical Resonance Imaging require additional computation time, recording and are much expensive \cite{anzivino2012fabian},\cite{kawahara2017brainnetcnn}. 
The basic concept of EEG is to measure the electrical activity of the brain with reference to time. The EEG signals are quite complex and can benefit from the deep learning algorithm in various practical applications \cite{luo2020biomarkers},\cite{phang2019multi} and,\cite{movahedi2017deep}. 
%\textcolor{red}{Provide Reference of below sentence}\textcolor{green}{done}
It is also been reported in various electrophysiological studies that abnormalities in gamma and theta EEG frequency band are directly related to memory deficiency in Sz patients\cite{phang2019multi}.
Therefore, EEG signals have been identified as a diagnostic tool and effective biomarker compared to others for the automatic detection of many neurological disorders due to their economical and non-invasive nature \cite{hornero2006variability},\cite{kindler2011resting}. Over the past few decades, many EEG-based deep learning and machine learning algorithms have been proposed for the effective detection of various neurological disorders such as Depression \cite{sharma66dephnn}, Parkinson’s \cite{oh2018deep}, epilepsy, Alzheimer’s \cite{hampel2010biomarkers} and Person identification \cite{wilaiprasitporn2019affective}. Moreover, several processes can also be implemented using EEG online systems due to the rapid growth in wearable EEG Devices.\\
The structure of the paper is organized as follows: Section II presents the background and contribution, respectively. Section III introduces the deep learning model for Sz detection. The methodology along with the datasets is explained in Section IV. The implementation are elaborated in section V. Section VI covers the results and discussion are carried out in Section VII. Finally, the conclusion is presented in section VIII.

\section{Background}
In recent years, researchers have demonstrated the potential of standard deep learning and machine learning techniques for various mental disorder \cite{yasin2021eeg,tadalagi2021autodep, shuai2017comprehensive}. The schizophrenia is considered as one of the 
 critical brain abnormalities  \cite{namazi2019fractal,ravi2016deep}. For example, many machine learning algorithms have been implemented for the automatic classification of Sz using imaging data, primarily based on neuroanatomical features and neuroimaging features \cite{olejarczyk2017graph}. 
%\textcolor{red}{place recheck this statement.It is strong statement} \textcolor{green}{sir, This statement is refering to the preveiously mentioned studies for refrence 26,27,28 as in next section EEG based discussion is given.}
All above mentioned studies were based on imaging data but have not considered EEG-derived features to explore brain abnormalities for detection of Sz disorder. EEG signals have shown promising results to detect brain abnormalities and are much more convenient in comparison to imaging data \cite{borisov2005analysis}. \\
Kim et al.\cite{kim2015diagnostic} presented a model with 62.2\% classification accuracy for Sz detection wherein EEG signals were extracted from 21 Gold cup electrodes with 10-20 international standard positioning. A study of vertical and horizontal eye movements was done. Thereafter, five frequency bands were selected using MATLAB and EEGLAB  toolboxes. The spectral power was calculated using Fast Fourier Transformation for each frequency band. After this, analysis of variance (ANOVA) and receiver operation curve (ROC) techniques were used to determine the diagnostic test performance. Dvey-Aharonet et al.\cite{dvey2015schizophrenia}, employed Time-Frequency Transformation on the EEG for Sz disorder wherein first the EEG was converted into images using Stockwell transformation. Then, the features were extracted and a classification procedure was implemented which obtained an accuracy between 92.0\% to 93.9\%. Johannesen et al.\cite{johannesen2016machine}, employed Brain Vision Analyser to analyze EEG, after which EEG was segmented through four stages of processing: pre-stimulus baseline, encoding, retention, and retrieval. The wrapper method was used for feature selection and two different SVM models were used for the classification of normal and Sz subjects. SVM Model 1 reported  84\% and SVM model 2 reported 87\% classification accuracy. Santos-Mayo et al.\cite{santos2016computer} acquired EEG with the help of Brain Vision equipment by 10-20 international standard. After pre-processing, 4 frequency domain and 16 time-domain features were extracted. After which, classification was done using SVM and Multilayer perceptron which results in 93.4\% and 92.23\% classification accuracy for Sz. Ibanez-Molina et al.\cite{ibanez2018eeg} presented a model wherein EEG data acquisition was done using Neuroscan SynAmps 32 channel amplifier. In this work, the moving window method was used for segmentation and Lempel-Ziv complexity was computed for every window. Then, the final LZC value was calculated by taking the average of the values obtained from every segment.
Jahmunah et al.\cite{jahmunah2019automated}
extracted many non-linear features, out of which 14 main features were identified based on their significance level. In this, various classifiers were employed but SVM with the Radial-basis function reported best performance with an accuracy of 92.91\% in comparison to other classifiers.\\
These studies were based on the traditional machine learning approach but in recent years, many researchers have demonstrated deep learning techniques for the evaluation of brain signals and mental health disorders. The main advantage of using deep learning compared to machine learning is that the model itself retrieves features from the data and feature engineering is not required.
Phang et al.\cite{phang2019classification} discriminated against the normal and Sz subjects using a deep learning approach CNN. In this, 1-D intricate network features and 2-D time and frequency features were extracted, which then fed into the CNN model for the classification process. This model reported accuracy of 93.05\%. Oh et al.\cite{oh2019deep}, implemented an 11 layered deep convolutional neural network model for the classification of Sz. This study proposed two separate models for subject-based and non-subject-based testing. The subject-based testing reported a classification accuracy of 81.26\% and the non-subject-based testing reported a classification accuracy of 98.07\%.\\
In our paper, an effective hybrid system based on deep learning methods is proposed for the automatic diagnosis of Sz disorder using EEG signals which is incorporated with IoMT in smart connected healthcare. The proposed hybrid model is a cascade of CNN and LSTM \cite{chandran2021eeg} wherein, EEG signals are initially processed and then fed to a deep neural network for classification.
The main contributions of this work are summarized as follows:\\
$\bullet$ The hybrid EEG-based Sz detection model for smart healthcare is developed with IoMT framework. It enables to provide feedback for necessary preventive actions with closed loop control system.\\
%$\bullet$ A fully automated deep learning model is proposed, wherein manual feature extraction is not required which can hamper the performance of the classification model.\\
$\bullet$ A fully automated deep learning model is proposed, wherein manual feature extraction is not required \cite{pancholi2020fast}. This model enables learning on both local features and long-term dependencies for EEGs. The proposed model reports the highest accuracy using EEG for the Sz dataset, which has not been reported earlier with any other biomarker for any Sz dataset.\\
$\bullet$ For the first time, examination is done using EEGs from various sets of limited electrodes and EEGs from different frequency bands for Sz detection.\\
$\bullet$ The proposed model has been designed and evaluated using various parametric settings for Sz detection. Windowing is used as a part of pre-processing to training time.\\
$\bullet$ The performance of the proposed model is evaluated using two different and larger datasets which justifies the model performance and reliability. The model has also been evaluated on various parametric settings with 5 electrodes only.

\section{Deep Learning Approach to EEG}
\begin{figure}[htbp]
\centerline{\includegraphics[scale=0.7]{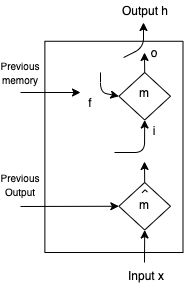}}
\caption{1-D convolutional operation}
\label{fig:1D}
\end{figure}

In recent years, there has been a  major boost in deep learning-based techniques for the classification of EEG signals \cite{sun2021hybrid}.  Researchers have explored both the spatial and temporal information out of EEG through deep learning architectures \cite{kokate2021classification}. 
 CNN models have shown remarkable results in extracting meaningful features or patterns from images. These CNN network has showed promising results when it is used as the primary block of any deep learning system for several applications. The convolution operation using kernels (small filter patches) is the primary feature of CNN. Figure \ref{fig:1D} shows the process of 1-D CNN. The left side of figure is a matrix of time-series data wherein it represents a kernel and convolution is done from top to bottom. 
The dimension of extracted feature is Mx1 after the convolution operation which is related to input data dimension, filter size and convolution step length. Such filters are capable of learning local features automatically, which can be mixed to create more complex patterns when multiple CNN layers along with the pooling layers are built together.  The features extracted from the CNN are majorly used as input to other networks which perform classification. 

\begin{figure}[htbp]
\begin{center}
\subfigure[CNN]{\label{fig:CNN1}\includegraphics[scale=0.45]{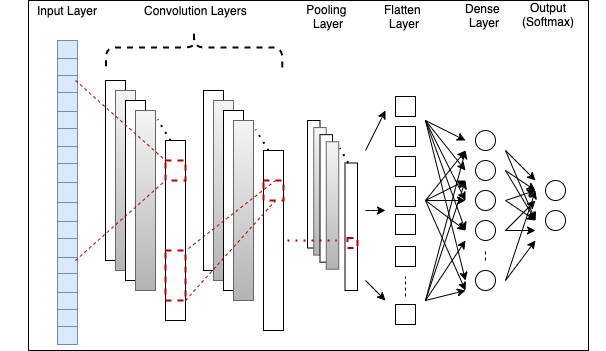}}
\subfigure[LSTM]{\label{fig:LSTM1}\includegraphics[scale=0.6]{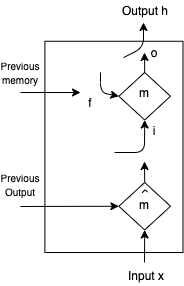}}
\end{center}
\caption{Basic (a) CNN and (b) LSTM structure and their operation}
\label{fig:CNN-LSTM}
\end{figure}

The main concern with the CNN is that it can not deal with the temporal information of any time series data. However, as the EEG signals are time-series signals, accessing temporal information can be much more advantageous. The overview of basic CNN and LSTM structures and their operation is shown in Figure \ref{fig:CNN-LSTM}.

LSTM has demonstrated promising results in dealing with any time-series data as EEG.  These networks are much better at handling temporal information because of their ability to record, memorize and discard the information depending on the nature of the system \cite{hochreiter1997long} \cite{shi2015convolutional}. Figure \ref{fig:LSTM1} shows the interconnection of an LSTM unit. Let's $x_{t}$ be the input to an LSTM unit at time step t and $h_{t}$ is the output. $h_{t}$ is a vector of outputs from all individual units $h_{t}^{k}$, where k represents the index of the LSTM cell. LSTM is primarily consisted of three gates; input gate ($i_{t}^{k}$,), output gate ($o_{t}^{k}$) and forget gate($f_{t}^{k}$). Each LSTM unit also has an additional memory cell which is denoted by $m_{t}^{k}$.  The input gate, output gate and forget gate are computed as follows:
\begin{equation}
i_{t}^{k}=Y^{k}(P_{i}x_{t}+Q_{i}h_{t-1}+R_{i}c_{t-1})
\end{equation}
\begin{equation}
o_{t}^{k}=Y^{k}(P_{o}x_{t}+Q_{o}h_{t-1}+R_{o}c_{t})
\end{equation}
\begin{equation}
f_{t}^{k}=Y^{k}(P_{f}x_{t}+Q_{f}h_{t-1}+R_{f}c_{t-1})
\end{equation}
Where $P_{i}$, $P_{o}$, $P_{f}$, $Q_{i}$, $Q_{o}$, and $Q_{f}$ are trainable weight matrices for updates. $R_{i}$, $R_{o}$, and $R_{f}$ are trainable diagonal matrices which keeps the memory parameters within each LSTM unit.  $Y_{k}$ takes the $k^{th}$ index, and passes it through a non-linear function as sigmoid.
The memory component gets updated with new component $\hat{c}_{t}^{k}$  using the equation given below and forgets the existing one:
\begin{equation}
c_{t}^{k}=f_{t}^{k}c_{t-1}^{k}+i_{t}^{k}\hat{c}_{t}^{k}
\end{equation}
And the new memory component can be calculated as follows:
\begin{equation}
\hat{c}_{t}^{k}=Z_{k}(P_{c}x_{t}+Q_{c}h_{t-1})
\end{equation}
 where $Z_{k}$ is a tanh non-linearity. From equation (4), it is clear that new memory component is dependent on forget and input gates. The final out from LSTM unit is computed as follows:
\begin{equation}
h_{t}^{k}=o_{t}^{k}\tanh(c_{t}^{k}) 
\end{equation}
\begin{figure*}[htbp]
\centerline{\includegraphics[scale=0.65]{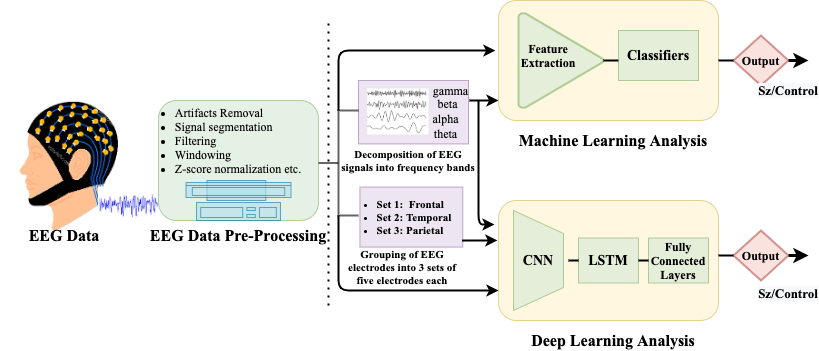}}
\caption{Overview of EEG-derived Sz detection using machine learning and deep learning methods}
\label{fig:arch}
\end{figure*}
Therefore, An efficient model can be implemented utilizing both CNN and LSTM to process EGG  (Time-series) for much more promising results.  In this study, a hybrid deep neural network is proposed which is a cascade of CNN followed by an LSTM network wherein, preceding CNN layers work as an automatic feature extractor for the latter LSTM network, which then classifies the Sz and Normal subjects. Figure \ref{fig:arch} shows an overview of the implemented Sz detector with the help of EEG using machine learning and deep learning algorithm \cite{pancholi2019time, mumtaz2018machine, wu2020detecting, pancholi2019electromyography}. The framework consists of two stages: EEG signal extraction and classification. Initially,  EEG has been acquired from the scalp using EEG Headset with or without any external stimulation to the subject. Then the preprocessed EEG is first applied to machine learning based model wherein hand crafted features are extracted \cite{pancholi2021intelligent} from EEG and then traditional classifier are used to identify Sz. 
The pre-processed EEG is also applied to deep learning model wherein CNN network extracts all the meaningful features from EEG and fed to the LSTM network. LSTM network performs sequence learning on the data coming from CNN which then helps to study the importance of each feature in conclusion making. Details of each process are described in the next section.
\section{Methodology}
In this section, the datasets along with pre-processing and experimental setup is explained in details. The experimental analysis and the performance evaluation of the proposed model is also described. Then, since the datasets are created to classify Sz and normal patients, the data partitioning methodology is illustrated. Finally, the proposed Hybrid deep learning model and its implementation is demonstrated.
\subsubsection{Datasets}
The models have been implemented and evaluated using two publically available datasets.
\paragraph{Dataset 1}
In this study, the EEG signals are acquired from 14 subjects with paranoid Sz (7 males and 7 females, with a mean age of 27.9 $\pm$ 3.3 and 28.3 $\pm$ 4.1 years, respectively). EEG signals of 14 healthy controls with similar gender proportion and age group were also collected by the same institute for the study with their consent. The dataset is publicly available which is provided by the Institute of Psychiatry and Neurology in Warsaw, Poland \cite{olejarczyk2017graph}. All the Sz subjects met ICD-10 benchmarks by the International Statistical Classification of Diseases and related health problems in the category F20.0 with at least one week washout period and a minimum age of 18 years.  Besides, Patients with severe neurological conditions such as epilepsy, Alzheimer’s and sleep arena or other medical consideration as pregnancy were excluded from the study. A multi-channel (19 channel) EEG data was obtained using the standard International 10-20 system. The EEG channels Fp1, Fp2, F7, F3, Fz, F4, F8, T3, C3, Cz, C4, T4, T5, P3, Pz, P4, T6, O1, and O2 with the reference electrode FCz were used. All the participants were asked to remain in a resting state with their eyes closed while collecting the EEG data of 15 min at the sampling rate of 250Hz.
The EEG dataset was pre-processed using the following steps:\\
$\bullet$ The EEG signals were divided into the non-overlapping segments of 25s without artifacts ( i.e. cardiac activity, eye movements, muscle contractions) for analysis.\\
$\bullet$ A bandpass filter from 4.0 to 45.0 Hz was applied to signals of each EEG channel which then further filtered into the following physiological frequency bands: theta (4-8Hz), alpha (8-15 Hz), beta (15-32Hz), gamma (32-45 Hz) and all bands (4-45 Hz).\\
$\bullet$ There were 19x6250 sampling points in each segment and a total of 1110 segments were used. These segments were further grouped into normal and Sz classes datasets.\\
$\bullet$ EEG signal of each segment was normalized using z score normalization before applying to models to scale signals in a standard range of values.
\paragraph{Dataset 2}
In this research, a publicly available EEG dataset was used to evaluate the performance of the proposed model in classifying healthy control and Sz subjects. The dataset was provided by the Laboratory for Neurophysiology and Neuro-Computer Interface at Lomonosov Moscow State  University \cite{kawahara2017brainnetcnn}.  The dataset consists of 84 subjects (39 healthy control and 45 Sz subjects), aged 11-14 years with a mean age of 12.3 years.  A multi-channel (16 channel) EEG was recorded from the participants of 1-minute duration at a sampling rate of 128 Hz as they remained in a resting state with eyes closed. The EEG channels F7, F3, F4, F8, T3, C3, Cz, C4, T4, T5, P3, Pz, P4, T6, O1, and O2 were used. All the Sz subjects were diagnosed at Mental Health Research Center (MHRC) and met the ICD-10 benchmark by the International Statistical Classification of Diseases and related health problems. Patients with any other neurological conditions were excluded from the research.
The EEG dataset was pre-processed using the following steps:\\
$\bullet$ Each file contained an EEG record of one subject wherein a column had EEG samples from all 16 channels. So initially reshaping of each file was done into 16x7680 sampling points. Then, these files were grouped to form a new dataset to classify normal and Sz subjects.\\
$\bullet$ A bandpass filter from 4.0 to 45.0 Hz was applied to signals of each EEG channel which then further filtered into the following physiological frequency bands: theta (4-8Hz), alpha (8-15 Hz), beta (15-32Hz), gamma (32-45 Hz) and all bands (4-45 Hz).\\
$\bullet$ These EEG signals were then normalized using z score normalization before applying to the models to scale signals in a standard range of values.
\subsubsection{Model Architecture}
As the dataset consists of EEG from Sz and normal subjects, an experiment was carried out to detect Sz disorder with the highest accuracy in deep learning-based applications. Two different approaches were implemented as shown in Figure \ref{fig:arch} : 1) Deep learning and 2) machine learning. EEG in all bands (4-45Hz) frequency range was used in the experiment. 
\paragraph{Deep-Learning}
Several ideas have been utilized for effective network architecture. The network is implemented and trained from scratch to learn interpretations of raw EEG signals with minimum numbers of trainable parameters. By extensive experiments, CNN, LSTM, and Hybrid CNN-LSTM models have been compared.  In CNN's, one or two-dimensional convolutional filters are trained to automatically extract features using multiple layered architectures. LSTMs consist of memory cells that are proved to be stable in sequential data modeling. On the other side, Hybrid CNN-LSTM is proved to be best for EEG data to extract important features wherein convolutional layers are followed by LSTM units to classify a deep learning model.
\begin{figure}[htbp]
\centerline{\includegraphics[scale=0.6]{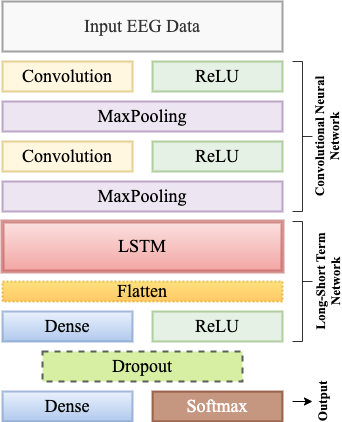}}
\caption{Internal architecture of proposed SzHNN model}
\label{fig:int}
\end{figure}
In this paper, A Hybrid CNN-LSTM model is adopted which has shown the superior performance for classification on EEG data (see section V). Initially, the input is applied to the first core block consisting of the CNN network to extract important features. Then, the CNN output is subsequently fed to the second core block consisting LSTM network. Finally, the output from the second core block is fed into a classification layer with a softmax activation function as shown in Figure \ref{fig:int}. The internal architecture and details of the proposed Hybrid model for EEG- based Sz detection are given in the Table \ref{tab:hybrid}. %Figure \ref{fig:LSTM}. 
This proposed model process raw signals while avoiding handcrafted features to identify relevant features at the early stage of psychosis. Let matrix $X\epsilon R^{TC}$ serves as input samples (for 19x6250 volumes) with paired labels $Y\epsilon R$, where $T$ denotes sampling points, $C$ denotes recorded EEG channels and $Y$ is 0 or 1 as label prediction. The proposed Hybrid model consists of Convolutional layers, Pooling layers, LSTM, and dense layers. \\
In this experiment, a 8 layered network with 2 core blocks (Convolutional and LSTM), two dense layers, and a softmax layer has been achieved. The first core block consists of two convolutional layers with 5 and 10 kernels with size [1x15] and [1x10] (experimentally evaluated), respectively. One max-pooling layer with kernel size 2 is used after each convolutional layer in the first core block. Non-linearity was added using the rectified linear unit. The second core block consists of a single LSTM unit whose dimension is set to 32 and one dense layer with a node number of 64. A dropout layer with the probability of 50\% was added after the second block for regularization. %The details of the proposed model are given in Table \ref{tab:hybrid}. 
\begin{table}[!htp]\centering
\caption{Configuration of the proposed SzHNN model}\label{tab:hybrid}
\scriptsize
\begin{tabular}{p{0.2em}ccp{5em}c}\toprule
\textbf{Layer} &\textbf{Type} &\textbf{Number} &\textbf{Size} &\textbf{Stride/Activation/Dropout}  \\\midrule
\multirow{2}{*}{1} &\multirow{2}{*}{Input} &\multirow{2}{*}{} &19x6250 (Dataset 1) &\multirow{2}{*}{} \\
& & &16x7680 (Dataset 2) & \\
2 &Convolutional &5 &[1x15] &1/ReLU/- \\
3 &Max Pooling & &[1x2] &2/-/- \\
4 &Convolutional &10 &[1x10] &1/ReLU/- \\
5 &Max Pooling & &[1x2] &2/-/-\\
6 &LSTM &32 &  & \\
7 &Dense &64 & & -/ReLU/0.5 \\
8 &Output &2 & & -/Softmax/-  \\
\bottomrule
\end{tabular}
\end{table}
%Initially, the input is applied to the first core block consisting of the CNN network to extract important features. Then, the CNN output is subsequently fed to the second core block consisting LSTM network. Finally, the output from the second core block is fed into a classification layer with a softmax activation function.
\paragraph{Machine-Learning}
Several machine learning methods have been explored through literature for effective implementation. After extensive study, Welch method is employed as a feature extractor in this implementation \cite{la2014human}. Welch is best known power spectral density estimation method. This divide the data into overlapped blocks forming periodogram after windowing. This further helps in reducing the variance of individual power after averaging the individual periodogram.
Finally, the SVM classifier is used for the classification of these hand crafted features \cite{pancholi2018portable}.
\section{Implementation}
For the implementation of the proposed model, several DL parameters such as the filter size, the number of channels, and the number of layers are evaluated experimentally. The classical back-propagation algorithm as a learning algorithm is used to tune the weights of the network, which can be seen by the improvement of model accuracy on the validation set \cite{islam2021emotion}. The performance improvement is also evaluated by observing intermediate normalization using batch normalization and dropout for regularization. The sparse categorical entropy function is used as a loss function for the training and performance evaluation of the proposed model. Adam optimizer with a learning rate of 0.0001 and decay rates of 0.0001 is used, respectively. The batch size is fixed on 32. The model is trained on over 100 epochs using default initialization parameters from Keras. Although, these parameters are found sufficient for parameter tuning is also explained in further sections. 
The proposed model is implemented on a workstation with an Intel i7 (2.30 GHz processor) and a 12 GB RAM DDR4 using Keras library. 
\subsubsection{Comparison of EEG-based Sz detection among EEGs from different frequency bands}
EEG is used to measure electrical activity across the scalp that occurs from the movement of billions of neural cells forming the human brain. Many studies were involved for the segmentation of such EEG into frequency bands for performance analysis. In this study, EEG is divided into five bands using butterworth bandpass filter: theta (4-8Hz), alpha (8-15 Hz), beta (15-32Hz), gamma (32-45 Hz), and all bands (4-45 Hz).  Thereafter, analysis is done to measure the model performance in Sz detection over various frequency bands.  For these experiments, CNN, CNN-LSTM, and SVM have been implemented for accuracy comparison.
\subsubsection{Comparison of EEG-based Sz detection among EEGs from sets of limited EEG electrodes}
\begin{figure}[htbp]
\begin{center}
\subfigure[Dataset 1]{\label{fig:dataset1}\includegraphics[scale=0.4]{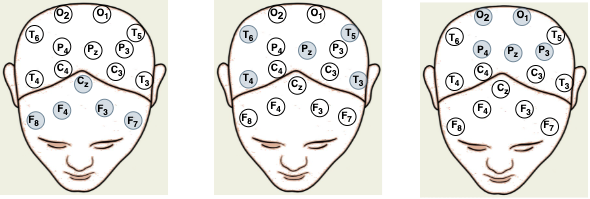}}\\
\subfigure[Dataset 2]{\label{fig:dataset2}\includegraphics[scale=0.4]{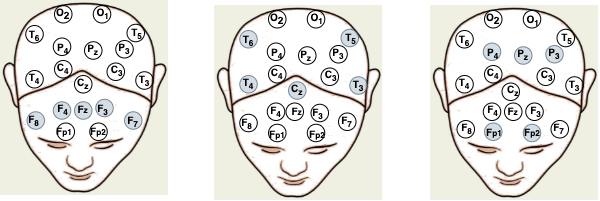}}
\end{center}
\caption{Performance of EEG-based Sz detection in terms of accuracy on different sets of EEG electrodes. Three sets of EEG electrodes were grouped from each part of scalp using both the datasets.}
\label{fig:datasets1}
\end{figure}

In this study, two different datasets with 19 and 16 EEG electrodes have been utilized for model implementation. If the number of EEG electrodes can be reduced, the system can become more practical and user-friendly. Therefore, A hypothesis has been done to find whether or not the number of EEG electrodes could be reduced from 19 or 16 to five while maintaining the accuracy. To investigate this, an experiment is conducted with 3 sets of EEG electrodes grouped from each part of the scalp as shown in the Figure \ref{fig:datasets1}, including temporal and parietal (TP), Frontal (F), and central and occipital (CO). Each set consists of 5 electrodes.
\subsubsection{Comparison of proposed Hybrid CNN-LSTM and other relevant approaches}
Initially, the proposed Hybrid model has been evaluated against the CNN and LSTM models. Details of both the models are given in Table \ref{tab:CNN1} and Table \ref{tab:LSTM}. In this study, the performance is evaluated in terms of accuracy and convergence speed as the tunning of CNN layers and LSTM units is done. The proposed model has also been compared with conventional machine learning algorithms such as SVM-based classifier using PSD as a feature and some existing relevant works proposed in literature.
\paragraph{Deep-learning Method}
To get an efficient model with a less limited number of CNN layers for cascading with LSTM units, an experiment is conducted with a varied number of CNN layers and LSTM units as shown in Table \ref{tab:PS1}, Table~\ref{tab:PS2} and Table~\ref{tab:PS3}. Subsequently, an efficient Hybrid Model is proposed with selected number of CNN layers and LSTM units. 
\begin{table}[!htp]\centering
\caption{Configuration of the implemented CNN model}\label{tab:CNN1}
\scriptsize
\begin{tabular}{p{0.2em}ccp{5em}c}\toprule
\textbf{Layer} &\textbf{Type} &\textbf{Number} &\textbf{Size} &\textbf{Stride/Activation/Dropout} \\\midrule
\multirow{2}{*}{1} &\multirow{2}{*}{Input} &\multirow{2}{*}{} &19x6250 (Dataset 1) &\multirow{2}{*}{} \\
& & &16x7680 (Dataset 2) & \\
2 &Convolutional &5 &[1x15] &1/ReLU/- \\
3 &Max Pooling & &[1x2] &2/-/- \\
4 &Convolutional &10 &[1x10] &1/ReLU/- \\
5 &Max Pooling & &[1x2] &2/-/- \\
6 &Convolutional &10 &[1x10] &1/ReLU/- \\
7 &Max Pooling & &[1x2] &2/-/- \\
8 &Dense &64 & &-/ReLU/0.5 \\
8 &Dense &32 & &-/ReLU/0.2 \\
9 &Dense &2 & &-/Softmax/- \\
\bottomrule
\end{tabular}
\end{table}
\begin{table}[!htp]\centering
\caption{Configuration of the implemented LSTM model}\label{tab:LSTM}
\scriptsize
\begin{tabular}{p{0.2em}ccp{5em}c}\toprule
\textbf{Layer} &\textbf{Type} &\textbf{Number} &\textbf{Size} &\textbf{Stride/Activation/Dropout} \\\midrule
\multirow{2}{*}{1} &\multirow{2}{*}{Input} &\multirow{2}{*}{} &19x6250 (Dataset 1) &\multirow{2}{*}{} \\
& & &16x7680 (Dataset 2) & \\
2 &LSTM &32 & & \\
3 &LSTM &64 & & \\
4 &Dense &32 & &-/ReLU/0.5 \\
5 &Dense &2 & &-/Softmax/- \\
\bottomrule
\end{tabular}
\end{table}
\begin{table}[!htp]\centering
\caption{Variation in number of filters for each layer while keeping LSTM units and kernel size same}\label{tab:PS1}
\scriptsize
\begin{tabular}{cccc}\toprule
\textbf{Number of Layers} &\textbf{CNN Filters} &\textbf{Kernel Size} &\textbf{LSTM Units} \\\midrule
1 &5 &15 &32 \\
2 &5, 10 &15, 10 &32 \\
3 &5, 10, 15 &15, 10, 5 &32 \\
\bottomrule
\end{tabular}
\end{table}
\begin{table}[!htp]\centering
\caption{Variation in kernel size for each layer while keeping the LSTM units and number of filters same }\label{tab:PS2}
\scriptsize
\begin{tabular}{cccc}\toprule
\textbf{Number of Layers} &\textbf{CNN Filters} &\textbf{Kernel Size} &\textbf{LSTM Units} \\\midrule
2 &5, 10 &5, 10 &32 \\
2 &5, 10 &10, 15 &32 \\
2 &5, 10 &15, 20 &32 \\
2 &5, 10 &15, 10 &32 \\
\bottomrule
\end{tabular}
\end{table}
\begin{table}[!htp]\centering
\caption{Variation in LSTM units for each layer while keeping the number of filters and kernel size same }\label{tab:PS3}
\scriptsize
\begin{tabular}{cccc}\toprule
\textbf{Number of Layers} &\textbf{CNN Filters} &\textbf{Kernel Size} &\textbf{LSTM Units} \\\midrule
2 &5, 10 &15, 10 &8 \\
2 &5, 10 &15, 10 &16 \\
2 &5, 10 &15, 10 &32 \\
\bottomrule
\end{tabular}
\end{table}
\paragraph{Machine-learning Method}
As mentioned earlier, SVM based classifier with PSD using the welch method is used as a feature implementation of the model. There have been several research papers which are based on schizophrenia detection using EEG with higher classification accuracy. Therefore, a machine learning model has been attempted to better comparison with previous models. 
To estimate PSD as a feature, the welch method with a 256 point hanning window and 50\% overlap is used. Each PSD feature consists of 19 electrodes for dataset 1 and 16 electrodes for dataset 2. Thereafter, classification is performed on transformed features followed by a logarithmic function to PSD features. Here, this method is used on all electrodes rather than choosing a subset for better performance comparison. 
\section{Results}
In this section, the performance of the proposed model is validated,  10 cross-validation is used in all experiments to obtain the accuracy. EEG-based Sz detection is compared amongst different datasets, frequency bands, set of EEG electrodes, other common structures, and competitive work to prove the superiority of the proposed model.
\subsubsection{Comparison of EEG-based Sz detection among EEG from different frequency bands}
\begin{figure}[htbp]
\begin{center}
\subfigure[Dataset 1]{\label{fig:dataset1freq}\includegraphics[scale=0.4]{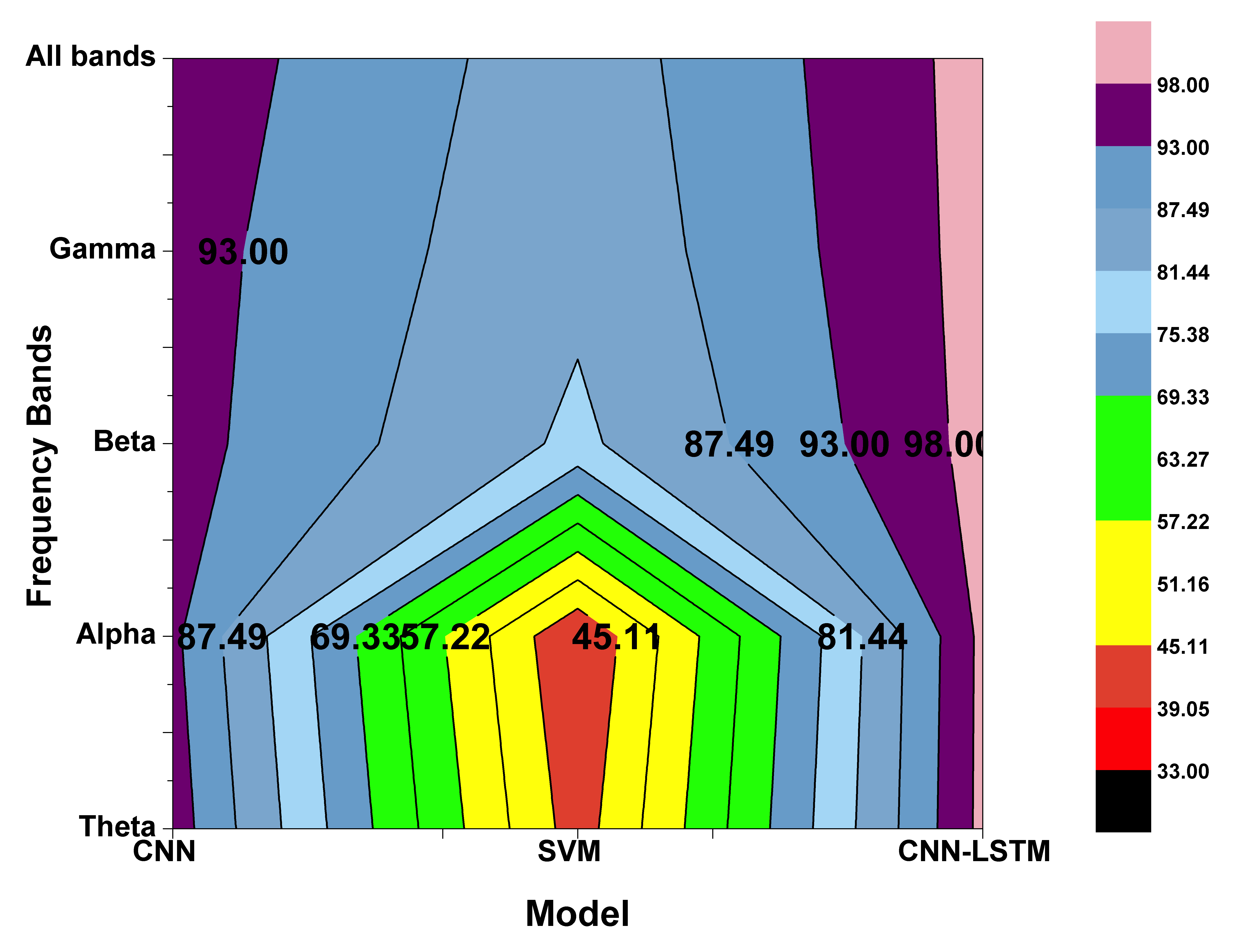}}\\
\subfigure[Dataset 2]{\label{fig:dataset2freq}\includegraphics[scale=0.06]{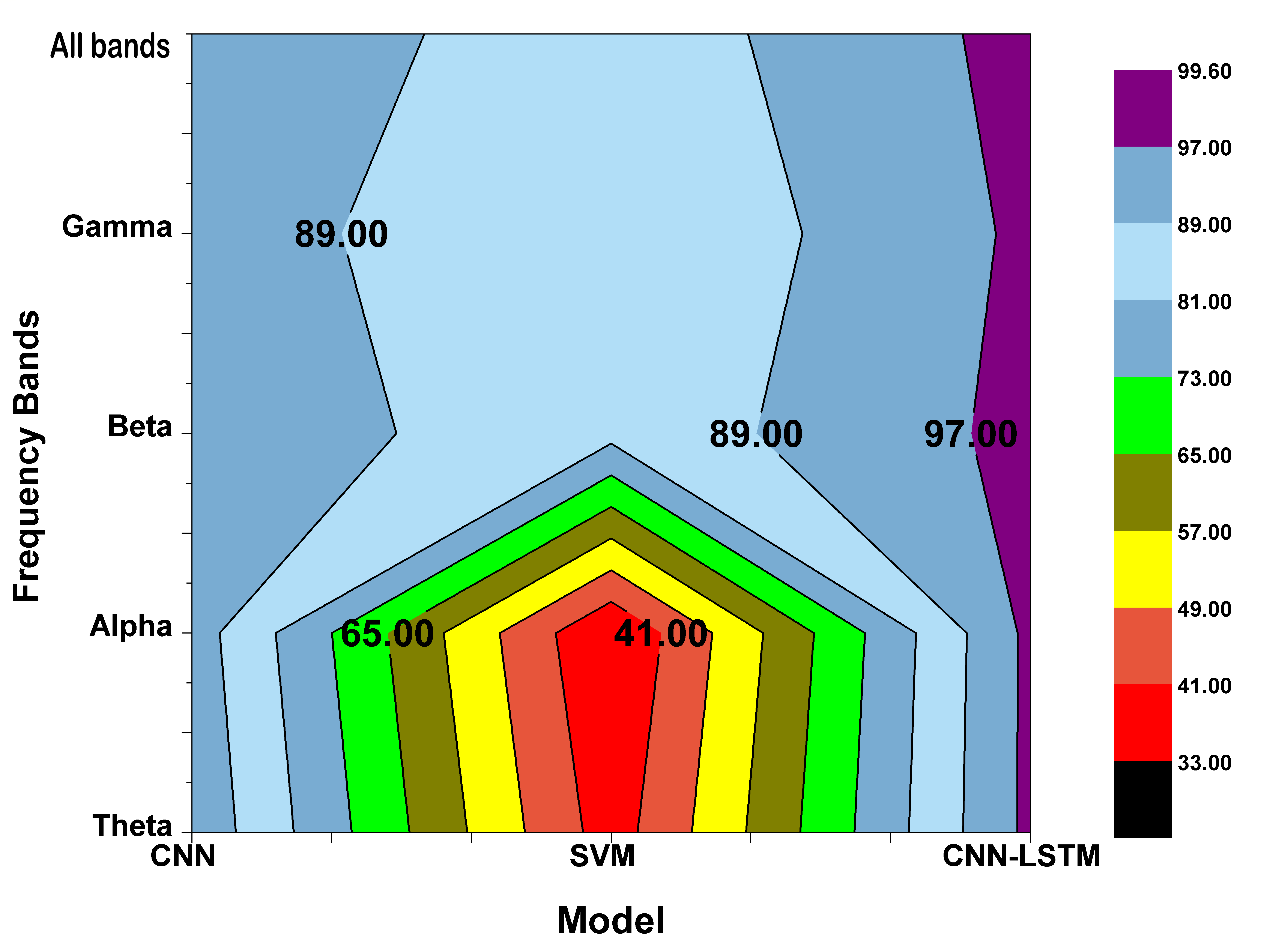}}
\end{center}
\caption{Comparison of the accuracies among different recognised approaches and among different EEG frequency bands. Theta(4-8Hz), Alpha(8-15Hz), Beta(15-32Hz), Gamma(32-45Hz) and All bands(4-45Hz)}
\label{fig:datasets}
\end{figure}

The comparison of the accuracy among different frequency bands using different approaches and datasets has been reported in Figure \ref{fig:datasets}. Statistical testing using two-factor ANOVA without replication along with standard t-test for follow-up analysis was also done for accuracy comparison \cite{pancholi2020advanced}. Statistical results with a 95\% level of confidence report that EEG from different frequency bands do not affect the performance of SZ detection using CNN and CNN-LSTM methods with (p>0.05). However, the SVM approach has showed that EEG from beta, gamma, and all bands provide significantly higher accuracy than the theta and alpha bands (p<0.05 for ANOVA testing with pairwise comparison) for both the datasets.
Moreover, for the accuracy comparison among various approaches, the statistical results have shown significant difference among CNN, CNN-LSTM, and SVM approaches (with p<0.05) in 95\% level of confidence. In pairwise comparison, the proposed Hybrid CNN-LSTM has significantly outperformed CNN and SVM in all available frequency bands and reported approx. 100\% accuracy using both
datasets which have shown excellent results as reported in Table (with p<0.05).
Here, all band frequency ranges have provided the best accuracy hence they have been used for all others remaining experiments.
\subsubsection{Comparison of EEG-Based Sz among EEG from different sets of electrodes}
Here, the Table \ref{tab:comp1} reported the accuracy comparison between three different sets of the electrode using CNN and CNN-LSTM approaches and two datasets.  Same statistical testing is used as mentioned above for this comparison. As mentioned earlier each electrode set consists of 5 electrodes. Statistical results reported that accuracy from set 1 including frontal electrodes significantly outperformed accuracies from the other two sets in 95\% level of confidence using both approaches and datasets (with p<0.05). However, the proposed hybrid CNN-LSTM has observed significantly higher accuracy till 96.10\% for dataset 1 and 91.03\% for dataset 2 than the other two approaches. Therefore, even after the reduction from 19 and 16 to five, the frontal electrodes have reported good accuracy so it can be explored in experiments of the same scenarios as this experiment.
\begin{table}[!htp]\centering
%\begin{adjustwidth}{-2.5 cm}{-2.5 cm}\centering\begin{threeparttable}[!htb]
\caption{Comparison of accuracies among different three sets of electrodes using CNN and CNN-LSTM approaches}\label{tab:comp1}
\scriptsize
\begin{tabular}{lccccc}\toprule
\multicolumn{5}{c}{\textbf{Accuracy (\%)}} \\\midrule
\textbf{} &\multicolumn{2}{c}{\textbf{CNN}} &\multicolumn{2}{c}{\textbf{CNN-LSTM}} \\
\textbf{} &\textbf{Dataset 1} &\textbf{Dataset2} &\textbf{Dataset 1} &\textbf{Dataset2} \\
\textbf{Set 1 (Frontal)} &84.68 &77.89 &96.10 &91.00 \\
\textbf{Set 2 (Temporal)} &74.23 &70.23 &89.00 &87.59 \\
\textbf{Set 3 (Parietal)} &59.23 &61.19 &63.10 &50.00 \\
\bottomrule
\end{tabular}
%\end{threeparttable}\end{adjustwidth}
\end{table}
\subsubsection{Comparison of accuracy results with various parameter settings}
The comparison of accuracy with variable CNN layers, kernel size, and fixed LSTM unit is presented in Table \ref{tab:compPS1}. The results have indicated that a CNN-LSTM setting consisting of two CNN layers with 5 and 10 filters of (1x15) and (1x10) kernel sizes, and one LSTM layer of 32 units reported the highest accuracy of 99.9 \% than the other two settings.
Furthermore, Table \ref{tab:compPS2} shows the accuracy comparison with fixed CNN layers and variable kernel sizes, and fixed LSTM unit. It is evident from the reported results that accuracy is decreasing if kernel size is being reduced. Now further experiment is done to see whether LSTM unit size can be decreased from 32 without affecting the accuracy. This can be seen in Table \ref{tab:compPS3}, which represents the accuracy comparison with smaller LSTM units while keeping both CNN layers and kernel sizes as the same. The results show that if smaller LSTM units reported the fewer units. Therefore, a single LSTM layer of 32 units is utilized in the proposed Hybrid model.
\begin{table}[!htp]\centering
\caption{Accuracy comparison with variation in number of filters for each layer while keeping LSTM units and kernel size same}\label{tab:compPS1}
\scriptsize
\begin{tabular}{cccc}\toprule
\textbf{CNN Filters} &\textbf{Kernel Size} &\textbf{LSTM Units} &\textbf{Accuracy (\%)}\\\midrule
5 &15 &32 &89.10 \\
5, 10 &15, 10 &32 &99.90 \\
5, 10, 15 &15, 10, 5 &32 &93.30 \\
\bottomrule
\end{tabular}
\end{table}
\begin{table}[!htp]\centering
\caption{Accuracy comparison with variation in kernel size for each layer while keeping the LSTM units and number of filters same}\label{tab:compPS2}
\scriptsize
\begin{tabular}{cccc}\toprule
\textbf{CNN Filters} &\textbf{Kernel Size} &\textbf{LSTM Units} &\textbf{Accuracy (\%)}\\\midrule
5, 10 &5, 10 &32 &93.60 \\
5, 10 &10, 15 &32 &97.10 \\
5, 10 &15, 20 &32 &89.20 \\
5, 10 &15, 10 &32 &99.90 \\
\bottomrule
\end{tabular}
\end{table}
\begin{table}[!htp]\centering
\caption{Accuracy comparison with variation in LSTM units for each layer while keeping the number of filters and kernel size same}\label{tab:compPS3}
\scriptsize
\begin{tabular}{cccc}\toprule
\textbf{CNN Filters} &\textbf{Kernel Size} &\textbf{LSTM Units} &\textbf{Accuracy (\%)}\\\midrule
5, 10 &15, 10 &8 &86.20 \\
5, 10 &15, 10 &16 &96.10 \\
5, 10 &15, 10 &32 &99.90 \\
\bottomrule
\end{tabular}
\end{table}
\subsubsection{Comparison of proposed model with other relevant implemented and existing approaches for EEG based Sz detection}
\begin{figure}[htbp]
\centerline{\includegraphics[scale=0.5]{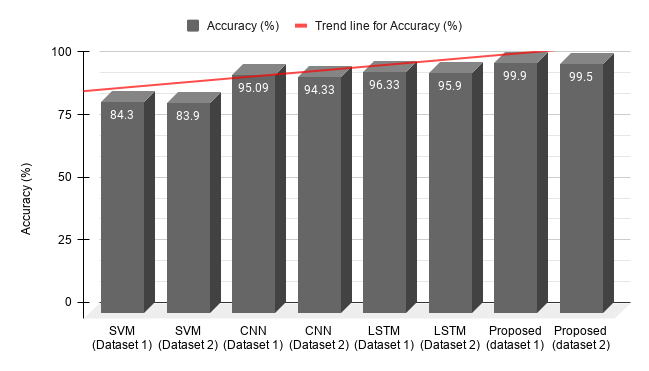}}
\caption{Acuuracies of compared methods}
\label{fig:comp chart}
\end{figure}

\begin{figure}[htbp]
%\begin{center}
\centering
\subfigure[Training accuracy by epoch for LSTM model]{\includegraphics[scale=0.4]{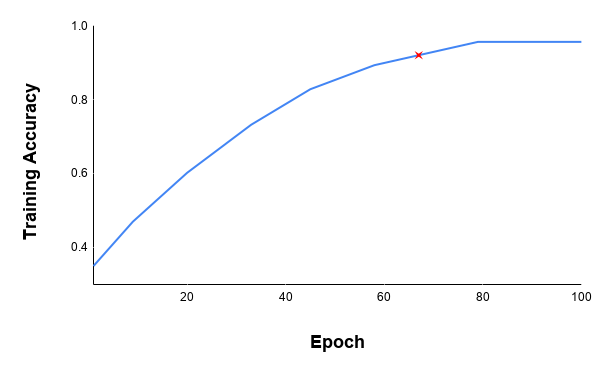}}\\
\subfigure[Training accuracy by epoch for CNN model]{\includegraphics[scale=0.4]{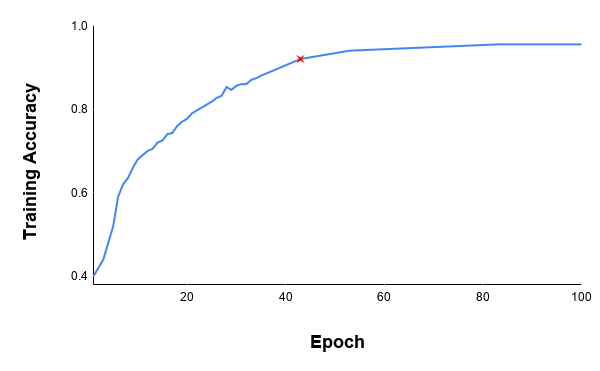} }\\
\subfigure[Training accuracy by epoch for proposed SzHNN model]{\includegraphics[scale=0.4]{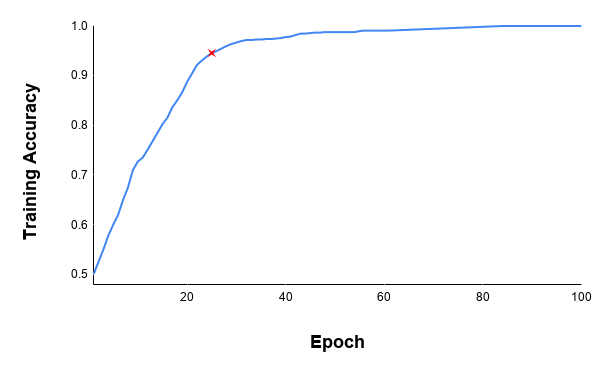} }
\caption{Compariosn of Training accuracy by epoch between CNN, LSTM and Proposed SzHNN model. Its shows the comvergence speed of proposed model is mcuh higher then other two implemented models}
\label{fig:structure}
\end{figure}

Table \ref{tab:comp final} represents the performance comparison of EEG-based Sz detection  using the proposed Hybrid CNN-LSTM with other approaches. 
Figure \ref{fig:comp chart} includes accuracy results from all deep learning and machine learning models implemented in this study on both datasets. Table  \ref{tab:comp final} incorporates accuracy results from some other relevant existing models using the same datasets from previous works. The proposed Hybrid CNN-LSTM (highlighted in bold) using 32 EEG electrodes reported the highest accuracy than others.  Figure \ref{fig:structure} shows the training speed comparison (in terms of training accuracy by epoch) from CNN only, LSTM only and Proposed SzHNN models. It is evident from Figure \ref{fig:structure} that training accuracy from SzHNN has increased much faster then other two models.
As a part of another experiment, the EEG signals were sent to the cloud too. These EEG signals are then processed and classification for the same is performed on the cloud itself, this results in an accuracy of 99.2\% which is comparable to the accuracy observed at the local server. 
\begin{table}[!htp]\centering
\caption{Comparison of accuracy among proposed, implemented conventional deep learning/machine learning approaches and the previous works \textbf{(Best results are highlighted)}.}
\label{tab:comp final}
\scriptsize
\begin{tabular}{p{5em}p{7em}p{4em}p{4em}p{4em}r}\toprule
\textbf{Year} &\textbf{Model} &\textbf{Number of Electrodes} &\textbf{Number of Participants} &\textbf{Accuracy (\%)} \\\midrule
2016\cite{johannesen2016machine}&Regression based SVM &64 &52 &84.00 \\
2017\cite{santos2016computer} &MISF-SVM &17 &47 &93.42 \\
2019\cite{jahmunah2019automated}$^{1}$ &RBF-SVM &19 &28 &92.91 \\
2019\cite{oh2019deep}$^{1}$ &CNN &19 &28 &81.26/98.07 \\
2019\cite{phang2019multi}$^{2}$ &MDC-CNN &16 &84 &91.69 \\
2019\cite{phang2019multi}$^{2}$ &MDC-SVM &16 &84 &90.37 \\
2020\cite{siuly2020computerized} &EMD-EBT &64 &81 &89.59 \\
2021\cite{chandran2021eeg}$^{1}$ &LSTM &19 &28 &99.00 \\
This paper$^{1}$$^{\#}$ &SVM &19 &28 &84.30 \\
This paper$^{2}$$^{\#}$ &SVM &16 &84 &83.90 \\
This paper$^{1}$$^{\#}$ &CNN &19 &28 &95.09 \\
This paper$^{2}$$^{\#}$ &CNN &16 &84 &94.33 \\
This paper$^{1}$$^{\#}$ &LSTM &19 &28 &96.33\\
This paper$^{2}$$^{\#}$ &LSTM &16 &84 &95.90 \\
Proposed Work$^{1}$ &CNN-LSTM &5 &28 &96.10 \\
Proposed Work$^{2}$ &CNN-LSTM &5 &84 &91.00 \\
\textbf{Proposed Work$^{1}$} &\textbf{CNN-LSTM} &\textbf{19} &\textbf{28} &\textbf{99.90} \\
\textbf{Proposed Work$^{2}$} &\textbf{CNN-LSTM} &\textbf{16} &\textbf{84} &\textbf{99.50} \\
\bottomrule
\\
$^{1}$Dataset 1 
$^{2}$Dataset 2&$^{\#}$Implemented conventional approaches
\end{tabular}
\end{table}
\section{Discussion}
In some recent years, researchers have explored the potential of several deep learning and machine learning methods for brain machine interface applications \cite{coogan2018brain, pancholi2019improved, sakhavi2018learning,pancholi2021novel}. They have been used extensively for the evaluation of mental illnesses. As discussed earlier, Sz is also a major mental disorder affecting people all around the world so several works have presented various EEG-based models to distinguish between SZ subjects and control subjects. EEG signals have demonstrated better performance as a biomarker for Sz detection due to their less complex nature in Sz subjects in comparison to control subjects. In this study, two kinds of issues have been highlighted, namely algorithmic and physical. The algorithmic issues referred to modeling and implementation of various deep learning (CNN only, LSTM only, CNN-LSTM) approaches for EEG-based Sz detection and it also reported the advantages of the proposed hybrid model over other related machine learning approaches. 
The physical issues indicate an effect of EEG capturing from different frequency bands and different sets of electrodes.  
In this paper, the majority of experiments have been conducted using two different EEG datasets. Regarding the algorithmic issues, machine learning algorithms using hand crafted features such as SVM with PSD did not perform well in obtaining good segregation between control and Sz subjects. So it is quite important to be aware of limitations and challenges while applying such algorithms to psychosis. On the contrary, the deep learning-based proposed hybrid CNN-LSTM algorithm demonstrated high performance in terms of accuracy in comparison to other deep learning models. Even with the fewer number of electrodes, proposed algorithms has reported the best accuracy among all existing and implemented algorithms. This information based on electrode location can be used in the interpretation of possible discriminative features.
As for the physical issues, from the reported results in the previous section, it is evident that the proposed deep learning-based models (CNN only, LSTM only, CNN-LSTM) could deal with all types of frequency bands and reach up to the classification accuracy of 100\%. On the other hand, machine learning-based models such as SVM using PSD as a feature did not perform well on all frequency bands and reported around 60\% mean accuracy. However, on specific frequency bands such as beta and gamma, accuracy using SVM was high up to 89.23\%. The performance of deep learning approach is least dependent on different frequency bands as these model doesn’t deal with hand crafted features and can capture inconspicuous features \cite{pancholi2021robust}. Similarly, it is also evident from the results that accuracy from a frontal set of electrodes are higher than other electrodes from different positions on the scalp. 
The proposed model is a crucial step in providing an automated tool for the early detection of Sz disorder. However, the model performance can be evaluated on a larger dataset for achieving ideal performance due to increased heterogeneity and different brain activity of participants.
\section{Conclusion}
This paper presents an experimental comparison among various machine and deep learning models for analyzing EEG based detection of Sz based on variable parametric settings. A novel hybrid CNN-LSTM model is proposed which consists of a combination of CNN for automatic feature extraction and LSTM for classification which reported a classification accuracy of 99.9\% with 32 electrodes and 96.10\% with 5 electrodes. The experiments are conducted using two different datasets namely dataset 1 of 19 subjects and dataset 2 of 16 subjects. The proposed model has significantly outperformed other relevant deep learning and machine learning algorithms on all available frequency bands even with a fewer number of electrodes. The proposed system has also been integrated with IoMT for smart connected healthcare wherein assessment and detection of Sz patients can be done remotely. However, the more robust model can be developed using a much larger dataset for Sz detection and proposed model can be tested in real-time healthcare environment in future.

% or


%%%%%%%%%%%%%%%%%%%%%%%%%%%%%%%%%%%%%%%%%%%%%%%%%%
\bibliographystyle{IEEEtran.bst}
\bibliography{paperbib1}

\pagebreak
%%%%%%%%%%%%%%%%%%%%%%%%%%%%%%%%%%%%%%%%%%%%
%%%%%%%%%%%%%% Authors' Bio
%%%%%%%%%%%%%%%%%%%%%%%%%%%%%%%%%%%%%%%%%%%%%%%%%%%%%%%
\section*{About the Authors}

\begin{IEEEbiography}
	[{\includegraphics[height=1in,keepaspectratio]{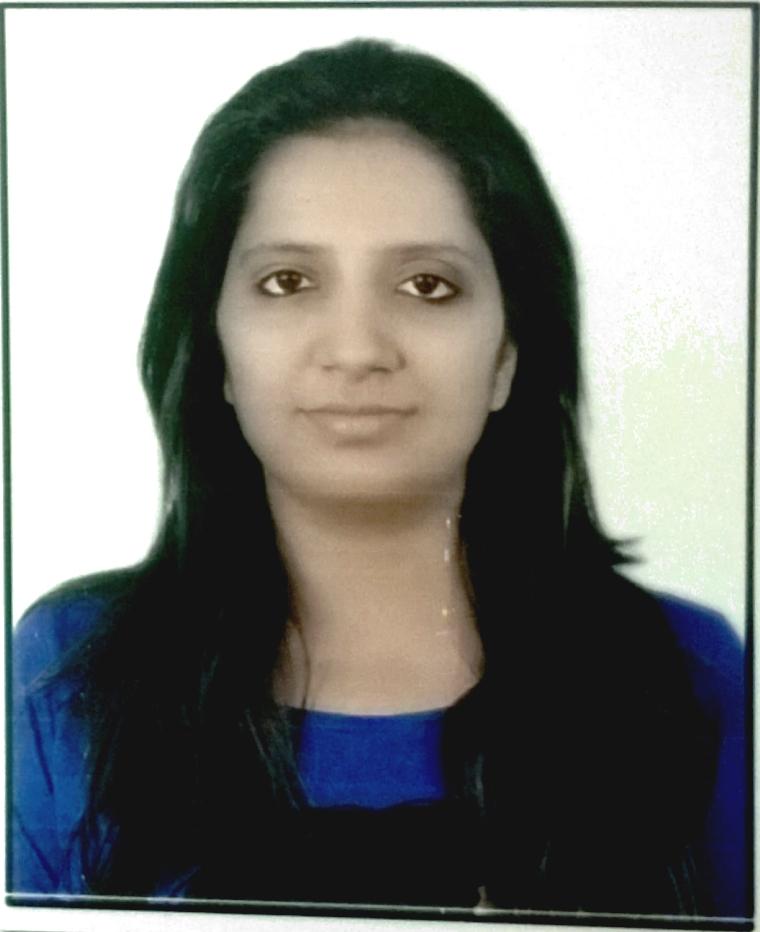}}] 
	{Geetanjali Sharma} (Member, IEEE) is working as an Assistant Professor in Electronics and Communication Department at MSIT (GGSIPU), New Delhi since last 10 years. She is also pursuing PhD from Electronics and Communication Department at MNIT Jaipur, Rajasthan. She has vast experience in the field of Low Power VLSI and has several research papers in reputed International Journals and Conferences.
\end{IEEEbiography}

\vspace{-0.8cm}

\begin{IEEEbiography}
   [{\includegraphics[height=0.8in,keepaspectratio]{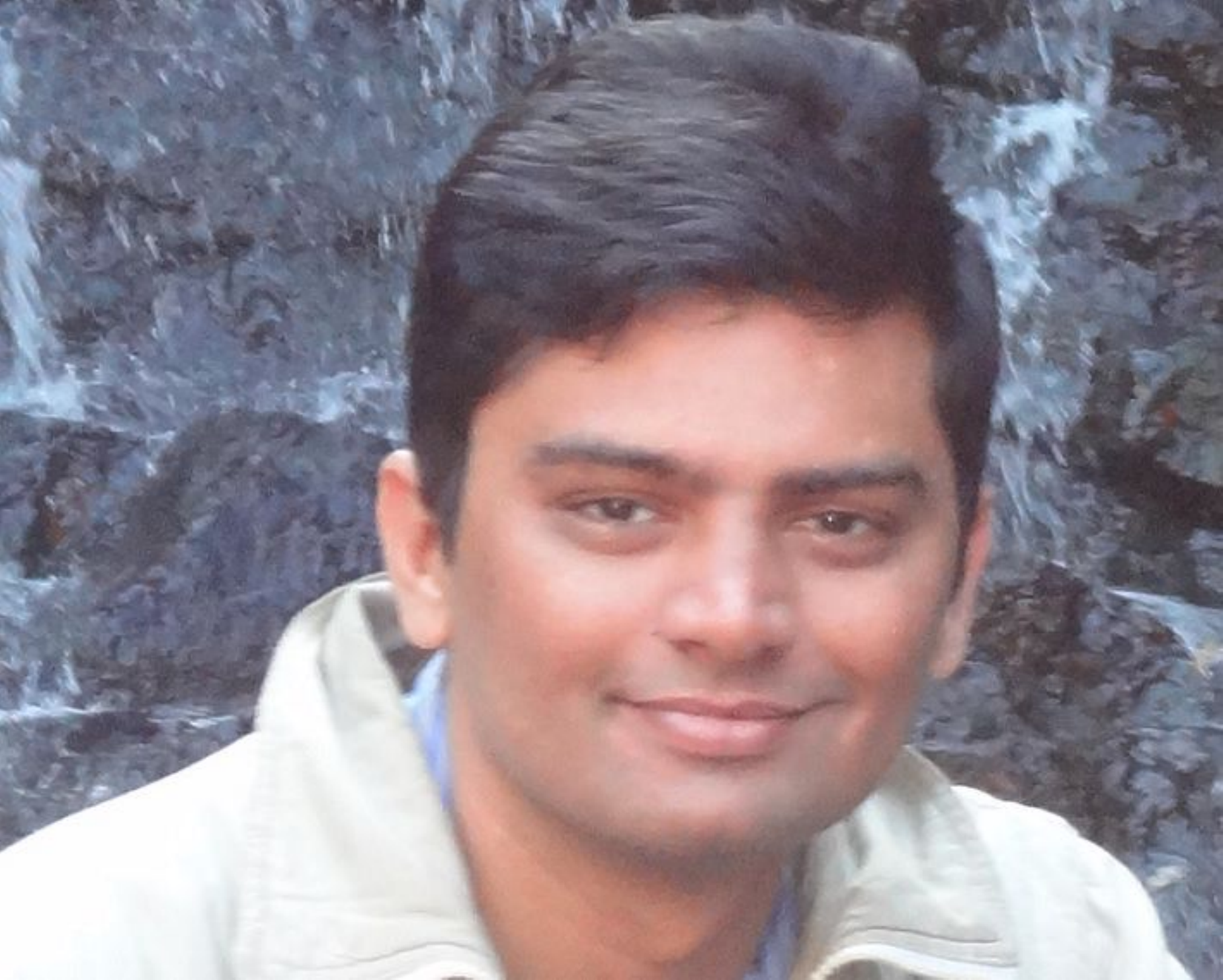}}] 
	{Amit M. Joshi} (Member, IEEE) received the Ph.D. degree from the NIT, Surat, India. He is currently an Assistant Professor at National Institute of Technology, Jaipur. His area of specialization is Biomedical signal processing, Smart healthcare, VLSI DSP Systems and embedded system design. He has also published papers in international peer reviewed journals with high impact factors. He has published six book chapters and also published more than 70 research articles in excellent peer reviewed international journals/conferences. He has worked as a reviewer of technical journals such as IEEE Transactions/ IEEE Access, Springer, Elsevier and also served as Technical Programme Committee member for IEEE conferences which are related to biomedical field. He also received honour of UGC Travel fellowship, the award of SERB DST Travel grant and CSIR fellowship to attend well known IEEE Conferences TENCON, ISCAS, MENACOMM etc across the world. He has served as session chair at various IEEE Conferences like TENCON -2016, iSES-2018, iSES-2019, ICCIC-14 etc. He has also supervised 19 PG Dissertations and 16 UG projects. 
He has completed supervision of 4 Ph.D thesis and six more research scholars are also working.
\end{IEEEbiography}

\end{document}